\documentstyle[amssymb,aps]{revtex}
\tightenlines 
\draft
\begin{document}
\title{Accuracy of equation-of-state formulations}
\author{Ronald E. Cohen, Oguz G\"{u}lseren, and Russell J. Hemley}
\address{Geophysical Laboratory and Center for High Pressure Research, Carnegie\\
Institution of Washington, 5251 Broad Branch Road, N.W., Washington, DC 20015}
\date{\today}
\maketitle

\begin{abstract}
The accuracy of equation-of-state formulations is compared for theoretical
total energies or experimental pressure-volume measurements for H$_{2}$, Ne,
Pt, and Ta. This spans the entire range of compression found for minerals and volatiles in the Earth.
The Vinet equation is found to be most accurate. The origin of
the behavior of different equation-of-state formulations is discussed. It is
shown that subtle phase transitions can be detected by examining the
residuals from an equation-of-state fit. A change in the electronic
structure of Ta is found at high pressures using this procedure, and a
possible new transition in H$_{2}$.
\end{abstract}

\section{Introduction}

In geophysics and high-pressure research, experimental or theoretical data
consisting of pressure, temperature, and volume triplets PVT are
parameterized to a functional form for ease of interpolation and
extrapolation. These equations of state are then used to compute phase
diagrams, or are used in geodynamic or shock compression modeling, etc. A
recent book by Anderson (1995) comprehensively reviews
equation-of-state formulations, and comprehensive reviews and comparisons of
equations of state are given by Hama and Suito (1996), Stacey et al. (1981), and
Duffy and Wang (1998). The primary purpose of this paper is to discuss the
reasons for the accuracy of the most-used formulation, based on the Birch
equation of state (Birch 1978) and compare with the Vinet equation of
state (Vinet et al. 1987). Jeanloz (1988) has previously shown that the Birch
equation of state and the Vinet equation of state can be similar up to
moderate compressions. Here we show that this breaks
down at high compressions, and for highly compressible materials the Vinet
equation of state is considerably more accurate. A logarithmic equation of
state (Poirer and Tarantola 1998) and another exponential equation of state
(Holzapfel 1996) are discussed and compared with the Vinet and Birch
equations. Finally, we show how subtle transitions can be detected by
deviations from an equation of state.

The above equations of state are appropriate for isothermal data. To include
thermal expansivity there are three main approaches that have been used. The
first is to fit isotherms, and then tabulate or fit the parameters $V_{0}$, $%
K_{0}$, and $K_{0}%
{\acute{}}%
$ as functions of temperature. Since experimental data are not always
collected along isotherms, that method is most amenable to analysis of
theoretical results. Secondly, one can assume Debye and Mie-Gruneisen
theory, and fit parameters for $\Theta _{D}$, $\gamma $, and $q$ to include
thermal effects. Thirdly, one can directly model the thermal pressure. See
Anderson (1995), Hama and Suito (1996), and Duffy and Wang (1998) for further
discussion of thermal equations of state.

In order to test equation-of-state formulations, it is important to study large compressions, because
most common equations of state will work reasonably over small compression ranges. Thus we choose to
study the very compressible materials, hydrogen and neon. We also consider tantalum and platinum,
which are useful pressure standards. 

\section{Isothermal or static equations of state}

The Birch equation of state (Birch 1978) is based on a series
expansion of pressure

\begin{equation}
P\left( f\right) =3K_{0}f\left( 1+2f\right) ^{\frac{5}{2}}\left[ 1+\frac{3}{2%
}\left( K_{0}^{\prime }-4\right) f+...\right]  \label{Birch}
\end{equation}
in terms of the Eulerian strain $f$, where

\begin{equation}
f=\frac{1}{2}[\left( \frac{V}{V_{0}}\right) ^{-\frac{2}{3}}-1].
\label{Eulerian}
\end{equation}
and $V_{0}$, $K_{0}$, and $K_{0}%
{\acute{}}%
$ \ are the zero pressure volume, bulk modulus, and bulk modulus pressure
derivative. Truncated as written it is called the ``third-order'' Birch
equation. The fourth-order Birch equation, which includes another term in $%
f^{2}$ and $K_{0}^{\prime\prime}$ is useful for theoretical equations of
state, but when applied to experimental data, the added parameter usually induces
severe correlations in parameters so that they lose physical significance.

The Vinet equation (Vinet et al. 1987) is derived from a scaled approximate
form for the energy: 
\begin{equation}
E\left( r\right) =-\Delta E\left( 1+a^{\ast }\right) \exp \left[ -a^{\ast }%
\right] ;\text{ }a^{\ast }=\frac{r-r_{0}}{l}  \label{universale}
\end{equation}
where $\Delta E$\ is the binding energy, and $r$\ is the length per
electron. This gives:

\begin{equation}
P\left( x\right) =3K_{0}\left( 1-x\right) x^{-2}\exp \left[ \frac{3}{2}%
\left( K_{0}^{\prime }-1\right) \left( 1-x\right) \right]  \label{Vinet}
\end{equation}
where $x=\left( \frac{V}{V_{0}}\right) ^{\frac{1}{3}}$. The energy can then
be expressed as

\begin{equation}
E=E_{0}+\frac{4K_{0}V_{0}}{(K_{0}^{\prime }-1)^{2}}-2V_{0}K_{0}(K_{0}^{%
\prime }-1)^{-2}(5+3K_{0}^{\prime }(x-1)-3x)\exp (-\frac{3}{2}(K_{0}^{\prime
}-1)(x-1))\text{.}  \label{vinete}
\end{equation}

The Holzapfel equation of state (Holzapfel 1996) is similarly given
by 
\begin{equation}
P(x)=3K_{0}x^{-5}(1-x)\exp \left[ (cx+c_{0})\left( 1-x\right) \right]
\label{Holzapfel2}
\end{equation}
where $c_{0}$ and $c$ are chosen to give $K^{\prime }$ and the limiting
Fermi gas behavior as $x\rightarrow 0$. If $c=0$, one gets a 3 parameter ($%
V_{0}$, $K_{0}$, and $K_{0}^{\prime }$) equation of state that behaves better at extreme compression 
(Hama and Suito 1996): 
\begin{eqnarray}
P(x) &=&3K_{0}x^{-5}(1-x)\exp \left[ c_{0}\left( 1-x\right) \right]
\label{Holzapfel} \\
&=&3K_{0}x^{-5}(1-x)\exp \left[ \frac{3}{2}(K_{0}^{\prime }-3)(1-x)\right] 
\nonumber
\end{eqnarray}

Finally we consider the logarithmic equation of state recently proposed by
Poirer and Tarantola (1998), derived similarly to the Birch equation, but where the
strain is defined as $\varepsilon =\ln \frac{l}{l_{0}}$. rather than as in
eq. \ref{Eulerian}, giving at third-order,

\begin{equation}
P=K_{0}\left[ \ln \frac{V_{0}}{V}+(\frac{K_{0}{\acute{}}-2}{2})(\ln \frac{V_{0}}{V})^{2}\right] . 
\label{log}
\end{equation}

Now we examine the application of these equations of state for several materials. First we consider
the extremely compressible behavior of hydrogen. Hemley et al. (1990) showed in analyses of x-ray
diffraction data for
hydrogen that the Vinet equation of state is considerably more accurate than
the Birch equation. The experimental equation of state, (room temperature
data is corrected to 4 K), is shown in Figure 1 along with equation-of-state
fits, and the fitted parameters and $\chi ^{2}$ are shown in Table \ref
{table:h2}. For most of the fits, $V_{0}$ was fixed at its known value, 23.0 cm%
$^{3}$/mole (Silvera 1980). The best fit is given by the Vinet equation of
state, with acceptable fits using Birch or Holzapfel, but the logarithmic
equation of state fails completely. When the parameter $V_{0}$ is relaxed
for the Vinet equation there is no significant change in the fit, and the
parameters become less well determined. Figure 1b shows the residuals of the
fits (the logarithmic residuals go off scale as shown). Interestingly, there
is a peak in the residuals at 40 GPa, which may be a subtle 
transition, or change in compression mechanism, previously undiscovered, or a small problem with a
subset of the
data. Note that this deviation is completely invisible on the P-V curve
itself. This illustrates the usefulness of having a good equation-of-state
formulation; even small deviations may indicate possible transitions that
should be examined more closely. Secondly, it illustrates that one only
expects equations of state to work well under compression with no underlying
transitions. If there are even subtle electronic or structural transitions,
these will affect the equation of state significantly.

One of the tests of an equation-of-state fit is the accuracy with which it
reproduces known zero-pressure parameters. To this end, high-pressure static
compression data can be compared with low-pressure elasticity measurements,
including data from ultrasonic experiments, Brillouin scattering, or in some
cases high-precision static compression data. Useful low-pressure static
compression data exist for hydrogen because of its very high
compressibility. Swenson and Anderson (1974) reported K$%
_{0T} = $ 0.17$\pm $0.06 GPa for n-H$_{2}$ from volumetric strain measurements to
2.5 GPa at 4.2 K. The results are in excellent agreement with those of
Wanner and Meyer (1973), who obtained K$_{0}=$ 0.174$\pm $%
0.010 GPa for single-crystal n-H$_{2}$ at 4.2 K using ultrasonic methods.
The latter number provides the best comparison with the present analysis
(e.g., for n-H$_{2}$). These results are also close to zero-pressure
Brillouin scattering results obtained at T=4 K for p-H$_{2}$ by Thomas et
al. (1978); the latter obtained K$_{0S}$ = 0.173$\pm $0.001,
with K$_{0T}$ $=$ 0.162 GPa (corrected for isothermal conditions).
Udovidchenko and Manzhelli (1970) performed
accurate static compression (volumetric) measurements on p-H$_{2}$ from 0 to
18 MPa (down to 6 K) and obtained K$_{0T}$ = 0.186$\pm $0.006 GPa. The neutron
diffraction study of p-H$_{2}$ to 2.5 GPa by Ishmaev et al. (Ishmaev et al. 1983) gave K$_{0T} = 
0.186\pm $0.03 and K$_{0T}^{\prime}$ $=$  6.33$\pm $0.2.
The Vinet $K_{0}$ is close to that determined directly, but the Birch and
Holzapfel $K_{0}$'s are way too high. The deviation in $K_{0,T}$ with the
Vinet equation may be due to being thrown off by the equation of state
glitch at 40
GPa, or may be due to insufficient flexibility in the Vinet
equation over this large compression range. In order to examine this possibility, we considered the
extended Vinet equation given by Moriarty (1995) and Vinet et al. (1989):

\begin{equation}
P=3K_{0}\left( 1-x\right) x^{-2}\exp \left[ \frac{3}{2}\left( K_{0}^{\prime
}-1\right) \left( 1-x\right) +\beta \left( 1-x\right) ^{2}+\gamma \left(
1-x\right) ^{3}...\right] \text{,}  \label{eq:exvinet}
\end{equation}
where $\beta =\frac{1}{24}\left( -19+18K_{0}^{\prime }+9K_{0}^{\prime
2}+36K_{0}K_{0}^{\prime \prime }\right) $ and fit the H$_{2}$ data varying $%
\beta $ as well as $V_{0}$, $K_{0}$, and $K_{0}^{\prime }$. As shown in
table \ref{table:h2}, $K_{0}=0.15\pm 0.02$, now in excellent agreement with
the ultrasonic bulk modulus. The peak in residuals survives this
higher-order fit. Other extensions to the Vinet equation are possible, and
it would seem more physical to add further terms to the prefactor of eq. \ref
{universale}, such as $b(a^{\ast })^{-c}$, rather than terms in the
exponential, but such extended equations of state will not be considered
further here.

Secondly, we consider another soft material, Ne, from fitting theoretical
Linearized Augmented Plane Wave (LAPW)(Singh 1994) local density
approximation (LDA)(Hedin and Lundqvist 1971) results. Figure 2 shows the fitted
equation of state, the residuals of the fit, and the resulting PV curves.
equation-of-state parameters are shown in Table \ref{table:ne}. Only the
Vinet and Birch equations are compared since the logarithmic equation of
state fails above for H$_{2}$, and the energy expression obtained from
integrating the Holzapfel PV equation is not closed form. The Vinet equation
is superior to the Birch equation again, and does an excellent job
simultaneously over the large energy scale of large compressions, and the
small energy scale near the minimum. We also show results for the
fourth-order Birch equation, which in spite of an additional parameter, is
still not as good as the Vinet equation. As more parameters are added, the
quality of the fit improves, but the correlations among individual
parameters increases.

Next we consider equations of state of two stiff metals, Ta and Pt (tables 
\ref{table:ta} and \ref{table:pt} and Figures 3 and 4). Energies were
computed using the LAPW method and the Generalized gradient approximation
[GGA-PBE, (Perdew et al. 1996)]. The Vinet equation fits better than the
Birch equation for Pt, but for Ta the Birch equation fits better. However,
examination of the residuals for Ta (Fig. 4c) shows that there is a \ peak
in the residuals at $\thicksim $10.5 \AA $^{3}$, and closer study
shows a change in the occupied bands (and thus the Fermi surface) around
that volume, where the t$_{2g}$ states at $\Gamma $ move below the Fermi
level, E$_{F}$, (Fig. 5) so the improved fit using Birch is a side effect of
this anomaly. Fitting only the data above 13 \AA $^{3}$ shows a greatly
improved fit, and a better fit for Vinet than Birch. This again illustrates
the utility of examining residuals from well founded equations of state such
as the Vinet equation in order to find subtle transitions.

\section{Discussion}

We have considered a range of materials from very compressible to more incompressible, covering the
entire range of compression of minerals and volatiles in the Earth. An equation of state reflects the
underlying interactions potential among
the ions and electrons that make up a crystal. Thus, a useful way to compare
different equations of state is to compare the implied assumptions about the
interaction potential. First we consider the Birch equation in some detail.
As normally expressed, it is not obvious that the Birch equation in terms of
Eulerian strain

\begin{equation}
E=\sum_{n=0}^{N}a_{n}f^{n}=\sum_{n=0}^{N}a_{n}\left[ \frac{1}{2}\left( \eta
^{-\frac{2}{3}}-1\right) \right] ^{n}\text{,}  \label{birche}
\end{equation}
where $\eta =\frac{V}{V_{0}}$, is actually a series for energy $E$ in $%
\left( \frac{V}{V_{0}}\right) ^{-\frac{2n}{3}}$. For example, if $N=3$

\begin{eqnarray}
E &=&a_{0}-\frac{1}{2}a_{1}+\frac{1}{4}a_{2}-\frac{1}{8}a_{3}+\left( \frac{1%
}{2}a_{1}-\frac{1}{2}a_{2}+\frac{3}{8}a_{3}\right) \eta ^{-\frac{2}{3}}
\label{birche3} \\
&&+\left( \frac{1}{4}a_{2}-\frac{3}{8}a_{3}\right) \eta ^{-\frac{4}{3}}+%
\frac{1}{8}a_{3}\eta ^{-2}  \nonumber
\end{eqnarray}
The $\frac{2}{3}$ powers arise from distance squared in finite strain
theory, so one can think of the Birch equation as a series in ``inverse
length squared.'' The reason only even terms in length are included is in
order to preserve rotational invariance in the strain energy expansion for
general strains. Thus, the Birch equation assumes that the underlying
potential can be represented as a series in $(1/r^{2n})$. The commonly used
third-order Birch equation (with parameters $V_{0}$, $K_{0}$, and $K_{0}%
{\acute{}}%
$) includes $n=1$,$2$, and $3$. Now, it is well known that polynomials can
behave poorly (wiggle) and extrapolate poorly (go crazy outside the range of
the fit), and this is in fact a problem with using the Birch equation to too
high an order. Note also that it is not a convergent series (unless the
coefficients obey $\frac{a_{n}}{2a_{n-1}}<f$, which is not expected in a
fit). Under compression, $\eta =V/V_{0}<1$ so that for larger $n$ the terms $%
\eta ^{-\frac{2n}{3}}$ blow up. The truncated series may still do well, as
it does up to moderate compressions, if the truncated series represents the
interatomic potential accurately. However, there is no fundamental reason to
expect $(1/r^{2n})$ to well represent interatomic interactions, so it is not
surprising that the Birch equation is not perfect. Indeed, the energy
expression from the Birch equation has a non-physical hump at expanded
volumes that then decays with increasing volume, whereas the Vinet and
Holzapfel equations converge smoothly to a constant at large volumes, being
consistent with physically based potentials. The logarithmic equation of
state unphysically diverges with expansion.

Another question that often arises with regard to the Birch equation is why $%
K_{0}%
{\acute{}}%
=4$ often works so well. Note that for moderate compressions (say 10\% or
less) higher order terms in eq. \ref{Birch} are quite small. For example, if 
$\eta =0.9$, then $f=0.036$, so that deviations from $K_{0}%
{\acute{}}%
=4$ are only reflected in a few percent or less of the pressure, and higher
order terms, with factor $f^{2}=0.0013$ are only affected at the 0.1\% level.

The Lagrangian strain 
\begin{equation}
\overline{\varepsilon }_{L}=\frac{1}{2}\left( \eta ^{\frac{2}{3}}-1\right) 
\text{.}  \label{lagrangianstrain}
\end{equation}
can be used instead of the Eulerian strain (Thomsen 1970), but the Lagrangian equation of state is not
satisfactory
at high pressures since it saturates with increasing density. The Lagrangian strain gives a series in
$E$

\begin{equation}
E=\sum_{n=0}^{N}a_{n}\overline{\varepsilon }_{L}^{n}=\sum_{n=0}^{N}a_{n}%
\left[ \frac{1}{2}\left( \eta ^{\frac{2}{3}}-1\right) \right]
^{n}=\sum_{n=0}^{N}b_{n}\eta ^{\frac{2n}{3}}  \label{Lagrangiane}
\end{equation}
which has the advantage that the series converges in $\eta ^{\frac{2n}{3}}$
for $\eta <1$, but since $\sum r^{2n}$ does not look like an interatomic
potential, it does not work well compared with the Eulerian strain.

The Hencky (logarithmic) strain (Poirer and Tarantola 1998) 
\begin{equation}
d\varepsilon _{H}=\frac{dl}{l};\varepsilon _{H}=\ln \frac{l}{l_{0}}
\label{henckystrain}
\end{equation}
shares a similar shortcoming. Though it looks reasonable as a measure of
finite strain the effective potential, $\sum \left( \ln r\right) ^{n}$ is
not very physical. The Hencky potential can be expanded to look like $%
-a(1/r)+b(1/r)^{(3K_{0}%
{\acute{}}%
-7)}$ (Poirer and Tarantola 1998). The attractive part is long-ranged and
ill-conditioned and the repulsive part is too soft.

In contrast, the Vinet equation is based on a potential at the outset (eq. 
\ref{universale}). This potential was introduced by Rydberg (1932), who used it not for an
equation of state (Stacey et al. 1981), but as a form for the {\it intra}molecular potential in
H$_{2}$
and other simple molecules for obtaining solutions to the anharmonic
Schrodinger equation for fitting and comparison to molecular vibrational
spectra. The Vinet equation of state works surprisingly well for a wide
range of types of materials, and for compressions of up to $\eta =0.1$, as
has been discussed in detail (Hama and Suito 1996; Vinet et al. 1987). Note
the sign errors in eq. 102 of Stacey et al. (1981).

One failure of the Vinet equation is that it does not give the proper
behavior at even greater compressions, and does not merge into 
electron gas (Thomas-Fermi), or quantum-statistical (Kalitkin and Kuz'mina 1972) behavior at extreme
compression. Holzapfel found
expressions that do have the proper limiting form, such as eqs. \ref
{Holzapfel} and \ref{Holzapfel2} above. The underlying energy expression,
however, of even the simplest Holzapfel equation is exceedingly complex

\begin{eqnarray}
E\left( x\right) &=&\frac{1}{8}x^{-5}K_{0}(9\exp \left[ \frac{3}{2}\left(
K_{0}^{\prime }-3\right) \left( 1-x\right) \right] \left(
-4+2(-5+3K_{0}^{\prime }\right) x  \label{Holzapfele} \\
&&+3\exp \left[ \frac{3}{2}\left( K_{0}^{\prime }-3\right) x\right] \left(
15-14K_{0}^{\prime }+3K_{0}^{\prime 2}\right) x^{2}%
\mathop{\rm Ei}%
\left[ -\frac{3}{2}\left( K_{0}^{\prime }-3\right) x\right] )\text{,} 
\nonumber
\end{eqnarray}
where

\begin{center}
\[
\mathop{\rm Ei}%
=-\int_{-z}^{\infty }\frac{e^{-t}}{t}dt\text{.} 
\]
\end{center}

The fact that the Holzapfel equation works less well than Vinet for H$_{2}$
and other materials (Hama and Suito 1996) up to almost 10-fold compression
suggests that this underlying potential does not accurately represent the
interaction potential in solids, though it has the correct limiting behavior
at extreme compressions. 

An equation of state that agrees with the quantum statistical model (Kalitkin and Kuz'mina 1972) at
extreme compression, which is obtained by assuming homogeneous compression of spherical atoms and
expanding the Hartree-Fock equations in density and dropping terms of order $\hbar^2$ and higher, was
developed by Hama and Suito (1996). This should become accurate at lower pressures than Thomas-Fermi.
We did not test this here, but in Hama and Suito's tests, it seems to work well from 0 to 10 fold
compression or more fore rare gases, simple metals, and ionic compounds. Further exploration of this
approach is warranted for extreme compressions. Their equation of state is:

\begin{equation}
P=3 K_0 x^{-5} (1-x)\exp[(\eta-3)(1-x)+(\xi-3/2)(1-x)^2]
\end{equation}
where $x=\left( \frac{V}{V_{0}}\right) ^{\frac{1}{3}}$, as above, and $\xi$ is determined from the
quantum statistical model (Kalitkin and Kuz'mina 1972).

Finally, in comparing equations of state, note that correlations among $%
V_{0}$, $K_{0}$ and $K_{0}^{\prime }$ are much lower for Vinet than Birch,
so parameters are better determined. In the Vinet equation, $K_{0}^{\prime }$
is inside exponential, and $K_{0}$ in the prefactor, whereas in the Birch
equation $K_{0}$ and $K_{0}^{\prime }$ appear as products. In any case, it
is very difficult to resolve $V_{0}$ from $V$-$P$ data, especially for a
high-pressure phase unstable at $P=0$. Data usually only extends to $P=0$ at
best, and it is thus difficult to obtain good values of $K_{0}$ and $%
K_{0}^{\prime }$ (if not for a robust equation-of-state formalism) since one
is trying to find first and second derivatives from one-sided data. This is
one advantage of theory; results are available at negative pressure and $%
V_{0}$ can be well constrained.

For strains less than 30\%, it probably doesn't matter what equation of
state you use, as was emphasized by Jeanloz (1988), but parameters
will still be better determined with the Vinet equation (see also Hemley et al. 1990). For large
strains,
the Vinet equation is best, and forms such as the Holzapfel equation are
required at extreme compressions.

\begin{acknowledgments}
This research was supported by NSF grants EAR-9418934 and EAR-9706624, the NSF Center for
High Pressure Research, and the Academic Strategic Alliances Program of the
Accelerated Strategic Computing Initiative (ASCI/ASAP) under subcontract no.
B341492 of DOE contract W-7405-ENG-48. Computations were performed on the
Cray J90/16-4069 at the Geophysical Laboratory, supported by NSF grant
EAR-9304624 and the Keck Foundation. We thank Orson Anderson, T. Duffy, and
D.G. Isaak and for helpful discussions.
\end{acknowledgments}


\newpage \squeezetable
\begin{table}[tbp]
\caption{Fitted equation-of-state parameters for H$_{2}$ pressure-volume
experimental data.}
\begin{tabular}{llllll}
& $\chi ^{2}$ & V$_{0}$ (cm$^{3}$/mole) & K$_{0}$ (GPa) & K$_{0}^{\prime }$
& K$_{0}$K$_{0}^{\prime \prime }$ \\ 
\tableline
Vinet & 0.1901 & 23.0 & 0.250$\pm $0.004 & 6.56$\pm $0.02 &  \\ 
Vinet & 0.1895 & 22.4$\pm $2.6 & 0.279$\pm $0.15 & 6.49$\pm $0.35 &  \\ 
Birch 3 & 0.6554 & 23.0 & 0.694$\pm $0.009 & 3.990$\pm $0.007 &  \\ 
Birch 4 & 0.2356 & 23.0 & 0.28$\pm $0.02 & 5.7$\pm $0.2 & -9.4$\pm $1.2 \\ 
Birch 4 & 0.4033 & 23.0 & 0.162 & 7.77$\pm $0.03 & -23.4$\pm $0.3 \\ 
Holzapfel & 0.4126 & 23.0 & 0.47$\pm $0.01 & 4.95$\pm $0.03 &  \\ 
Logarithmic & 13.98 & 23.0 & 0.08$\pm $0.36 & 64.3$\pm $292.4 &  \\ 
Ext. Vinet & 0.1623 & 23.0 & 0.15$\pm $0.02 & 8.0$\pm $0.4 & --21.$\pm $2.
\end{tabular}
\label{table:h2}
\end{table}

\begin{table}[tbp]
\caption{Fitted equation-of-state parameters for Ne LAPW (LDA) energy
(Ryd, 1 Ryd=13.605 eV))-volume results.}
\label{table:ne}
\begin{tabular}{llll}
& Vinet & Birch 3 & Birch 4 \\ 
\tableline$\chi ^{2}$ & 1.02$\times $10$^{-9}$ & 1.35$\times $10$^{-7}$ & 
3.03$\times $10$^{-9}$ \\ 
V$_{0}$ (cm$^{3}$/mole) & 8.641$\pm $0.008 & 8.598$\pm $0.074 & 8.592$\pm $%
0.013 \\ 
K$_{0}$ (GPa) & 8.94$\pm $0.06 & 11.2$\pm $0.8 & 9.24$\pm $0.18 \\ 
K$_{0}^{\prime }$ & 7.192$\pm $0.009 & 6.07$\pm $0.14 & 7.51$\pm $0.11 \\ 
K$_{0}K_{0}^{\prime \prime }$ & (-16.0)\tablenotemark[1] & (-10.2)%
\tablenotemark[1] & -21.1$\pm $0.98 \\ 
E$_{0}$ (Ryd) & -256.7593$\pm $0.00005 & -256.7658$\pm $0.0002 & -256.7654$%
\pm $0.00003
\end{tabular}
\tablenotetext[1]{Not varied, result of formulation.}
\end{table}

\begin{table}[tbp]
\caption{Fitted equation-of-state parameters for Pt LAPW (GGA-PBE) energy
(Ryd)-volume results (V$>$10\AA $^{3}$).}
\label{table:pt}
\begin{tabular}{llll}
& Vinet & Birch & exp.\tablenotemark[1] \\ 
\tableline$\chi ^{2}$ & 1.47$\times $10$^{-7}$ & 3.51$\times $10$^{-7}$ & 
\\ 
V$_{0}$ (cm$^{3}$/mole) & 9.448$\pm $0.007 & 9.445$\pm $0.011 & 9.09 \\ 
K$_{0}$ (GPa) & 248.9$\pm $0.7 & 238.6$\pm $1.1 & 278 \\ 
K$_{0}^{\prime }$ & 5.43$\pm $0.02 & 5.474$\pm $0.03 & 5.6 \\ 
E$_{0}$ (Ryd) & -36893.5648$\pm $0.0002 & -36893.5641$\pm $0.0002 & 
\end{tabular}
\tablenotetext[1]{(Holmes et al. 1989)}
\end{table}

\begin{table}[tbp]
\caption{Fitted equation-of-state parameters for Ta LAPW (GGA-PBE) energy
(Ryd)-volume results.}
\label{table:ta}
\begin{tabular}{lllllll}
 & Vinet & Birch & Vinet (V$>$13\AA $^{3}$) & Birch (V$%
> $13\AA $^{3}$) & exp. \tablenotemark[1] \\ 
\tableline $\chi ^{2}$ & 9.32$\times $10$^{-7}$ & 2.71$\times $10$^{-7}$ & 
8.08$\times $10$^{-9}$ & 1.31$\times $10$^{-8}$ &  \\ 
V$_{0}$ (cm$^{3}$/mole) & 11.026$\pm $0.009 & 11.058$\pm $0.005 & 11.057$\pm 
$0.002 & 11.062$\pm $0.03 & 10.865 \\ 
K$_{0}$ (GPa) & 188.$\pm $1.8 & 188.$\pm $0.9 & 192.3$\pm $0.2 & 190.$\pm $%
0.3 & 195$\pm$5 \\
K$_{0}$%
\'{}%
& 4.08$\pm $0.04 & 3.81$\pm $0.01 & 3.82$\pm $0.02 & 3.77$\pm $0.02 & 3.4$%
\pm $0.1 \\ 
E$_{0}$ (Ryd) & -31252.3337$\pm $\ 0.0004 & -31252.3339$\pm $0.0002 & 
-31252.33425$\pm $0.00004 & -31252.3342$\pm $0.00005 &  
\end{tabular}
\tablenotetext[1]{(Cynn and Yoo 1999)} \end{table}

\section{Figure Captions}

Figure 1. Equation-of-state of hydrogen. (a) Experimental P vs. V. and
equation-of-state fits. (b) Residuals of equation-of-state fits. Note that
the logarithmic equation-of-state residuals go off scale.

Figure 2. Equation of state of neon from LAPW total energy computations
within the LDA. (a) Energy versus volume and third-order Birch and Vinet
fits. The inset (inset area is shaded in the larger figure) shows the that
the Birch equation does not match the data at low pressures, whereas the
Vinet works at both low and high compressions. (b) Pressure volume relations
from the fits shown in (a) as well as a fourth-order Birch fit. In spite of
the fact that all fits look good to the eye at high pressures in figure (a),
the derived pressures differ by over 150 GPa at extreme compression. (c)
Residuals for the fits. Note the large and systematic deviations in the
third-order Birch fit. The fourth-order Birch is almost as good as Vinet,
but with one additional fitting parameter.

Figure 3. Static equation of state for Pt to extreme compressions computed
using LAPW and GGA. (a) Energy versus volume and Vinet, third- and
fourth-order Birch fits. The ``low P'' Vinet fit is to volumes greater than
10 \AA $^{3}$. (b) Pressure versus volume from the fits. The Vinet and
fourth-order Birch give similar results over the extreme pressure range. A
Vinet fit to a smaller pressure range gives better constrained equation of
state parameters and smaller residuals, but the pressures are very close to
those computed using the large data range. (c) Residuals of the Vinet fits,
and the lower pressure third-order Birch fit. Note that there are no
systematic deviations.

Figure 4. Static equation of state for Ta computed using LAPW and GGA. (a)
Energy versus volume. Note that to the eye both fits appear perfect. (b)
Pressure versus volume. In spite of the apparently excellent fits in (a),
the pressure is significantly different under extreme compression. (c)
Residuals of the fits. An electronic transition was hidden under the
apparently smooth equation of state. There is a peak in the residuals at
10.5 \AA $^{3}$. When data is fit at V$>$13\AA $^{3}$, the fit greatly
improves and the residuals are much smaller.

Figure 5. Band structures of Ta at (a) V=12.66 \AA $^{3}$ (5 GPa) and (b)
V=9.3 \AA $^{3}$ (460 GPa). There is a major change in the occupied states
as the t$_{2g}$ states move below the Fermi level E$_{F}$ at high pressures,
crossing at a volume of 11.67 \AA $^{3}$ (200 GPa). This change is
responsible for a glitch in the equation of state.

\end{document}